\documentstyle[12pt]{article}

\begin{document}


\hoffset = -1truecm
\voffset = -2truecm

\title{\bf
Plane Wave Limits and T-Duality 
}

\author{
{\bf
R. G{\" u}ven }\\
\normalsize The Abdus Salam International Centre for Theoretical Physics, Trieste
34100,
{\bf Italy}\\
{\normalsize and}\\
\normalsize Department of Mathematics, Bo\u{g}azi\c{c}i University, Bebek,
\.{I}stanbul 80815
{\bf Turkey}
\thanks{Permanent Address}}

\maketitle

\begin{abstract}
 The Penrose limit is generalized to show that, any leading
order solution of the low-energy field equations in any one of the five
string theories has a plane wave solution as a limit. This limiting
procedure takes into account all the massless fields that may arise
and commutes with the T-duality so that any dual solution has again a
plane
wave limit. The scaling rules used in the limit are unique and stem from the scaling
property of the D = 11 supergravity action.  
Although the leading order
dual solutions need not be exact or supersymmetric, their plane wave limits always
preserve some portion of the Poincar\'e supersymmetry and solve the relevant field
equations in all powers of the string tension parameter. Further properties of the 
limiting procedure are discussed.

\end{abstract}

\newpage

In general relativity there is a remarkably simple argument, due to
Penrose \cite{kn:pen}, which shows that any spacetime has a plane wave
as a limit. This universal property of the plane wave spacetimes can be
proven in two steps. For this purpose one first
chooses, on
an arbitrary Lorentzian spacetime, a coordinate gauge in a neighborhood of a
null geodesic. The coordinate patch chosen in this manner turns out to be
governed by the conjugate points of the null geodesic. One then utilizes the
homogeneity property of the Einstein-Hilbert action under the constant
scalings of the 
metric and blows up the neighborhood of the geodesic through the limit.
The endpoint of the
procedure
is always a plane wave spacetime which satisfies the Einstein equations if the
initial spacetime does.
 
    In this paper  we wish to study how the Penrose limit is generalized in string
theories and how it behaves under the T-duality. We shall work within the
framework of the low-energy effective field theories and take into account all
possible massless bosonic fields that may arise, including both the Neveu-Schwarz
(NS-NS) 
and the Ramond (R-R) sectors as
well as the Yang-Mills (YM) fields. It will be seen that the two inputs of the
Penrose limit both have natural generalizations in string theories. The gauge choice 
employed in the limit must be generalized to incorporate the antisymmetric tensor
fields and this can be done in a unified manner, applicable to all string theories .
Secondly, the different scaling behaviors that must be imposed on the
massless fields
to produce the plane waves
turn out to be rooted in the scaling property of the D=11 supergravity
action. Assuming that the D=10 spacetime possesses a spacelike isometry without
fixed points, but otherwise is arbitrary, we next consider the effect of
T-duality on the
conjugate points. We find that the conjugate points are left invariant by
the duality transformations. From this
observation and the invariance of the gauge conditions used, it follows that
T-duality commutes with the Penrose limit. Therefore,
starting from an arbitrary, leading order solution and its T-dual in any one of the
five string theories, one gets in the limit a plane wave solution together with its
dual plane wave in the dual theory.
The fact that the plane wave solutions of type I and heterotic string theories
can be obtained
through
the Penrose limit  was noted in \cite{kn:guv} and a related procedure was employed
in \cite{kn:tsy} to generate new, exact solutions possesing only the NS-NS fields.
Within the NS-NS sector, it is also
known that plane waves  constitute a T- duality invariant family when the isometries
that correspond to the translations along the wave fronts are 
gauged \cite{kn:kl}. Some  other interesting aspects of the string theory plane
waves
can be found in \cite{kn:ama}-\cite{kn:dev} and certain plane wave solutions of the
Type II theories are reported in \cite{kn:spin}, \cite{kn:seng}. 

     It is well known that the leading order terms of the low-energy Lagrangians of
the type II string theories are of the from
\begin{equation}
{\mathcal{L}}={\mathcal{L}}_{NS}+{\mathcal{L}}_R+{\mathcal{L}}_{CS},
\end{equation}
where the NS-NS sectors are described by the ten-form
\begin{equation}
{\mathcal{L}}_{NS}={\frac{1}{2 \kappa_{10}^2}}{ e^{-2\phi}} [ -R*1 + 4d \phi \wedge
*d \phi - \frac{1}{2} H \wedge *H ],
\end{equation}
and for the R-R sector of the IIA theory one has
\begin{equation}
{\mathcal{L}}_R={\frac{1}{4 \kappa_{10}^2}} [ F_2 \wedge *F_2 + F_4 \wedge *F_4 ],
\end{equation}
whereas for the IIB theory
\begin{equation}
{\mathcal{L}}_R={-\frac{1}{4 \kappa_{10}^2}} [ F_1\wedge *F_1 + F_3
\wedge *F_3 + {\frac{1}{2}} F_5 \wedge *F_5 ].
\end{equation}
Here $R$ is the D=10 scalar curvature in the string frame, $\phi$ is the
dilaton,
$H$ is the NS-NS
three-form: $H=dB$ and a subscript on a R-R field denotes the degree of that form.
Our spacetime conventions and the Hodge dual * are described in the appendix. 
The IIA  R-R field strengths are even degree forms which are defined in terms of
the odd degree potentials $A_p$  by
\begin{equation}
F_2 = dA_1, \hspace{1in} F_4 = dA_3 + A_1 \wedge H,
\end{equation} 
whereas in the IIB theory the even degree potentials $A_p$ give rise to the odd
degree R-R field strengths:
\begin{eqnarray}
F_1 = dA_0, \hspace{1in} F_3 =dA_2 + B \wedge dA_0, \nonumber\\
F_5 = dA_4 - {\frac{1}{2}} A_2 \wedge dB + {\frac{1}{2}} B \wedge dA_2 +
{\frac{1}{2}} B \wedge B \wedge dA_0.
\end{eqnarray}
The Chern-Simons terms ${\mathcal{L}_{CS}}$ do not affect the present discussion
and are ignored. In the IIB theory the field equations that follow from (1)
are in harmony with the self-duality of the five-form field strength:
\begin{equation}
F_5 = *F_5,
\end{equation} 
but (7) must be imposed as an additional field equation. For the IIA theory
(1) can be completely derived from the bosonic sector of the D=11 supergravity
Lagrangian 
\begin{equation}
{\hat{\mathcal{L}}} ={\frac{1}{2\kappa_{11}^2}} [{\hat{R}}\hat{*} 1 -
{\frac{1}{2}} \hat{F} \wedge \hat{*} \hat{F} 
-{\frac{1}{6}} \hat{F} \wedge \hat{F} \wedge \hat{A} ],
\end{equation}
by employing the standard Kaluza-Klein (KK) reduction. In (8)
$\hat{R}$ is the scalar curvature of the D=11 metric
$\hat{g}_{\hat{\mu}\hat{\nu}}$ and $ \hat{F} =d \hat{A}$
is the four-form field.

     For both the type I and the heterotic strings 
the leading order terms of the low- energy Lagrangians can be written as
$ {\mathcal{L}}={{\mathcal{L}}_1} + {{\mathcal{L}}_2} $, where $ {\mathcal{L}}_1 $
has
the
same form as $ {\mathcal{L}}_{NS} $ provided the Chern-Simons three-form of the YM 
field is included in the definition of $ H $, and ${{\mathcal{L}}_2} $ stands for
the
YM kinetic term with the appropriate gauge group and  dilaton coupling
\cite{kn:pol}.
\newcommand{\be}{\begin{equation}}
\newcommand{\ee}{\end{equation}}

        Consider now in the framework of the above Lagrangians the
massless fields of any one of the five string theories. Let us first introduce a coordinate system $ \{Y^+, Y^-,
Y^A\} $ with $A=1, \ldots ,8$ on the D=10 spacetime $ M_{10} $ so that the string
frame metric
takes the
form 
\be
ds^2= 2dY^+ [ dY^- +  \alpha dY^+ + \beta_A dY^A ] - C_{AB} dY^A dY^B,
\ee
where the metric functions $\alpha, \beta_A$, $ C_{AB} $ are in general functions of
all the
coordinates and $ C_{AB} $  is a $ 8 \times 8 $ positive definite symmetric matrix.
Such a
coordinate system can always be introduced in a neighborhood of a portion of a
null geodesic provided this portion contains no conjugate
points \cite{kn:rose}. These
coordinates have the propery that a null geodesic congruence is singled out in
which each geodesic is given by $ Y^+ , Y^A = const. $
with $ Y^+ $ labelling the different geodesics and $ Y^- $ is
an affine parameter along these geodesics. The coordinate system is valid
as long as a conjugate point is not encountered and breaks down at the
nearest
conjugate point where $ det (C_{AB}) = 0 $.

   In the NS-NS sector $( g_{\mu \nu}, \phi, B )$ one must choose a gauge
not only for the D=10 metric $ g_{\mu \nu} $ but also for $B$. Let the
components of $ B $ be labelled as
 \be B = B_{+-} dY^+ \wedge dY^- + B_{+A} dY^+ \wedge dY^A + B_{-A} dY^-
\wedge dY^A + 1/2
B_{AB} dY^A \wedge dY^B.
\ee
Then the appropriate gauge condition for $ B $ turns out to be 
\be
B_{-A} = 0,
\ee 
and this can always be imposed in the chosen neighborhood by using the gauge freedom
$B
\rightarrow B + d \chi$ with a suitable one-form $\chi$.
Each potential $A_p$ in a R-R sector also
enjoys a similar gauge freedom: $ A_p \rightarrow A_p + d \Lambda_{p-1} + \ldots$,
involving a $(p-1)$-form $ \Lambda_{p-1} $ and possible compensators that depend on 
$\chi$ or $H$ and whenever a R-R sector is present,
one must also arrange the gauge so that
\be
{A^{(p)}}_{- B_{1} \ldots B_{p-1}}= 0,
\ee
holds for each potential. Here we are denoting the components of $ A_p $ 
by $ A^{(p)}_{{\mu_1} \ldots{\mu_{p}}} $ in an expansion similar to (10) and $p \geq
1$.
In the cases where a YM field must be
taken into account, the same
applies to its Lie algebra-valued potential one-form $ {\mathcal A }$: 
\be
{\mathcal A}_- =0,
\ee
which means that we are working in the YM gauge: $ {\mathcal A}= {{\mathcal A}_+}
dY^+ +
{\mathcal A}_B dY^B $. This completes the first step of limiting procedure because
all the
gauges are now appropriately chosen. Notice that no restriction is made
on the dependence of the fields on the coordinates.

    The second step of the procedure starts by rescaling the coordinates chosen on
the
neighborhood. Let $ \Omega > 0 $ be a real number and
introduce $ \{ U, V, X^A \} $ satisfying \be Y^- = U, \hspace{.5in} Y^+ =
{\Omega^2} V, \hspace{.5in} Y^A = {\Omega}X^A.  \ee When the coordinate
basis one-forms are written in terms of $ dU, dV, dX^A $ in (9)  and (10)
but the components are not transformed, this rescaling gives us a
one-parameter family of fields $ ({g_{\mu \nu}}(\Omega), \phi(\Omega),
B(\Omega) ) $ and the same applies to the R-R as well as the YM fields
that are present. It is useful to view these as fields on a one-parameter
family of spacetimes $ {{M_{10}}}(\Omega) $. This allows one to interpret
$\Omega$ as a scalar field on an associated D=11 manifold possessing a
degenerate metric and a boundary \cite{kn:ger}. The boundary of the D=11
manifold is located at $\Omega =0$ which is the limit of interest. Before
approaching this boundary let us introduce on $ {M_{10}}(\Omega)$ new
fields
that are distinguished by overbars and are related to the old ones by \be
{{\bar{g}}_{\mu \nu}}(\Omega) = {\Omega ^{-2}} {g_{\mu \nu}}(\Omega), \ee
\be \bar{\phi}(\Omega) = \phi (\Omega), \ee \be \bar{B}(\Omega) = {\Omega
^{-2}} B(\Omega).  \ee For the R-R and the YM fields the scaling rules are
\be \bar{A}_p(\Omega) = {\Omega^ {-p}}A_p(\Omega), \ee \be \bar{\mathcal
A}(\Omega) = {\Omega^ {-1}}{\mathcal A}(\Omega), \ee so that each potential is
scaled according to its form degree. Allowing now $\Omega \rightarrow 0$,
the overbarred fields become in the limit:  \be ds^2 = 2dU dV - C_{AB}(U)
dX^A dX^B, \ee \be \phi = \phi(U), \ee \be B = \frac{1}{2}B_{KL}(U)dX^K
\wedge dX^L + gauge, \ee where we have dropped the overbars for notational
convenience. Notice that when $\Omega \rightarrow 0$ all the
functions appearing in each field depend only on the coordinate $U$ as a
consequence of (14). This is, of course, valid also for the components of the R-R
and YM fields. According to (5),(6) and (17),(18) a (p+1)-form R-R field
strength is scaled as \be {\bar{F}}_{(p+1)} = {\Omega^{-p}}F_{(p+1)}, \ee
and in the limit the scaled fields take the forms \be A_p = \frac{1}{p !}
A_{ K_1 \ldots K_p}(U) dX^{K_1} \wedge \ldots \wedge dX^{K_p} + gauge, \ee \be
F_{(p+1)} =
\frac{1}{p !} F_{- A_1 \ldots A_p}(U) dU \wedge dX^{A_1} \wedge \ldots
\wedge dX^{A_p}.  \ee Denoting the YM field strength by $\mathcal F $, one
also gets in the same limit 
\be
{ \mathcal A} = {{\mathcal A}_K} (U) dX^K + gauge,
\hspace{.5in} {\mathcal F} = {{\mathcal F}_{- K}}(U) dU \wedge dX^K. 
 \ee

What have been obtained by this limiting procedure are the general
representations of
the plane
wave fields in the Rosen coordinates. These coordinates have the virtue of
displaying the isometries but become singular at $det( C_{AB}(U)) =0$.
This 
can be remedied \cite{kn:gib} by transforming all the fields to the
harmonic coordinates $ \{u, v, x^A \}$:
\be
U = u, \hspace{.4in}V = v - \frac{1}{4} {\dot{C}}_{AB}(U) {Q^A}_ K (U) {Q^B}_ L (U) 
x^K
x^L,\hspace{.4in} X^A ={ Q^A}_ B (U) x^B,
\ee
which covers the whole of the plane wave manifold. Here and in the sequel a dot over
a quantity denotes differentation
with respect to its argument. The matrix ${Q^A}_ B$ is such that
\be
C_{KL} {Q^K}_ A {Q^L}_ B = \delta _{AB},\hspace{.5in}
C_{KL}[{\dot{Q}^K}_ {A}{Q^L}_  B - {Q^K}_ {A} {\dot{Q}^L}_ B ] = 0,
\ee
where $\delta_{AB}$ is the D=8 Kronecker symbol. Defining an $8 \times 8$ 
martix $h_{AB}(u)$ by
\be
 h_{AB} = - [ \dot{C}_{KL}{\dot{Q}^L}_  B  +  C_{KL}{\ddot{Q}^L}_  B ] {Q^K}_ A
\ee
the
spacetime line element (20) takes the standard form
\be
ds^2 = 2 dudv - h_{AB}(u) x^A x^B du^2 - \delta_{AB}dx^A dx^B.
\ee
In the harmonic coordinates the field strengths (but not the potentials) retain
their forms:
\be
\phi = \phi(u),\hspace{.3in} H = \frac{1}{2}H_{uAB}(u) du \wedge dx^A
\wedge
dx^B, \ee \be F_{(p+1)} = \frac{1}{p
!}F_{u
A_1 \ldots A_p}(u) du \wedge dx^{A_1} \wedge \ldots \wedge dx^{A_p},\ee
\be
{\mathcal F} = {\mathcal F}_{uA}(u) du \wedge dx^A.
\ee

It can be checked that
(15)-(19) which led us to the plane waves are the unique scaling rules that
produce
finite, non-zero field strengts. Remarkably, these rules also ensure that the D=10
Lagrangians transform
homogeneously:
\be
{\bar {\mathcal L}}(\Omega) = {\Omega^{-8}}{\mathcal L}(\Omega),
\ee
and it is possible to absorb $\Omega$ into the definition of the coupling
constant:
$\kappa_{10}^2 = {\Omega^8}{\overline {\kappa}}_{10}^2$.  A similar behavior is
encountered in the D=2 $ \sigma$-model Lagrangian $
{\mathcal L}_{\sigma}$ for the NS-NS fields. As was noted in \cite{kn:tsy} for a class of 
fields , $ {\bar{\mathcal L}}_\sigma {(\bar{\alpha '})} = {{\mathcal
L}_\sigma}{(\alpha ')}$ if one defines $ \alpha ' = {\Omega ^2}
{\bar{\alpha} '}$ where $\alpha ' $ is the string tension parameter. When viewed
from the
D=11 supergravity framework for the IIA teory,
one can see that (15)-(19) are precisely the D=10 consequences of the well known
scaling behavior of (8): \be \bar{\hat{\mathcal L}} = {\Omega^{-9}} {\hat{\mathcal
L}}, \ee under the transformations \be
{\bar{\hat{g}}}_{\hat{\mu} \hat{\nu}} = \Omega^{-2} {\hat{g}}_{\hat{\mu}
\hat{\nu}}, \hspace{.5in} \bar{\hat{A}} = \Omega^{-3} \hat{A}, \ee
and as we shall see below, the IIB theory scaling rules can then be
deduced via T-duality.

 An important consequence of (34) is that, if the fields
were chosen intially
to satisfy the relevant field equations, then (30)-(33) will again be a solution
of the same equations after the limit. In all such cases all the field equations
but one will be trivially satisfied by the plane waves and the remaining equation
will always be a
condition on the trace of $ h_{AB}(u) $, relating it to the other field strengths.
For the heterotic strings its precise form can be found in \cite{kn:guv} and for IIA
and IIB theories this equation will be displayed after we consider the T-duality.

   We have thus seen that any leading order solution in any of the five string
theories goes over to a
plane
wave solution in the limit. The plane wave family itself is closed under this
procedure because, the limit of a plane wave is always a plane wave
\cite{kn:ger}. The
limiting procedure, of course, makes no reference
to a symmetry of  the spacetime and (9) need not have any Killing vectors. Suppose
now
we assume that the initial $M_{10}$
admits 
a spacelike Killing vector $K^ \mu $ which has no fixed points. In such a situation
one would like to know whether the Penrose limit can be tried simultaneously on
a given solution and its T-dual and whether the limit of the
dual solution is also a plane wave. Since T-duality can even lead to a topology
change, it is not clear from the outset that the limit can be applied also to a dual
solution to get another plane wave. To proceed further one needs to understand the
dual
patch and see whether the same type of gauges
can be implemented on the dual fields. If the gauge conditions are preserved, then
clearly duality will commute with the Penrose limit and the dual of a plane wave
will always be a plane wave.

 Therefore, let us start by considering the NS-NS sector and assume  
that the action of $K^ \mu $ on the fields is specified in the standard
manner \cite{kn:roc},
\cite{kn:ag}. The T-duality transformations of the NS-NS fields \cite{kn:bush}
 take a simple form when the fields are decomposed relative to $K^\mu$ in
a
KK fashion \cite{kn:wel}. Let us introduce ${\lambda ^2} =
-K_\mu K^\mu$ and denote by $y$ the Killing coordinate: $Y^A = \{Y^j, y \}$,  $j =
1, \ldots ,7$. 
 Then the metric (9) can be decomposed as
\be
ds_{(10)}^2 = ds^2 - \lambda^2 (dy + \omega )^2,
\ee
where $ds^2$ is the D=9  KK metric and $\omega =\omega_+ dY^+ +\omega_j dY^j$ is the 
twist potential. Notice
that due to our gauge choice in (9), 
 $\omega_- = 0$ and therefore, the twist potential obeys a gauge
condition which is in perfect
harmony 
with
(11)-(13). We next define the one-form
$b
=
B_{y+}dY^+
+ B_{yj}dY^j$ and write
\be
B^{(10)} = B + dy \wedge b.
\ee
From now on, whenever there is a need to distinguish the D=10 fields from their D=9
descendants, our labelling of the D = 10 fields will be as in (37) or (38) with
either a subscript or a superscript. In terms of the
D=9 fields defined above it can be easily seen that T-duality leaves the KK metric
invariant and acts on the remaining NS-NS fields as
\be  
\tilde{\lambda} = \lambda ^{-1}, \hspace{.4in} \tilde{\omega} = b, \hspace{.4in}
\tilde{b} = \omega, \hspace{.4in} \tilde{B} = B +
b \wedge \omega, \hspace{.4in} \tilde{\phi} = \phi - \ln{\lambda},
\ee
where tilde denotes the T-dual of a field.

     Using these transformation rules and keeping in mind that the dual manifold
$\tilde{M}_{10}$ possesses a new Killing coordinate $\tilde{y}$ : ${\tilde{Y}}^A =
\{ Y^j, \tilde{y} \}$, 
it is useful to note first that, in string theory, a gauge choice for the axion
potential
must
necessarily
accompany the gauge choice for the metric. This is forced upon us by T-duality 
because, unless
the gauge for 
$B^{(10)}$ is chosen as in (11), the dual D=10 metric does not have the form of
(9) which is suitable for the limit.
More precisely, among the components of (11) it is the vanishing of  $B_{y-} \equiv 
b_{-}$ 
which
ensures the form invariance of the dual metric. Since the dual metric has the same
form in our gauge, it remains to see how the dual patch is related to the original
one. This is also
necessary because,
although the KK metric is left invariant, duality maps $ C_{AB}$ to a new matrix
$\tilde{C}_{AB}$ that contains contributions of $ \omega$ and $b$ and the locations
of the dual conjugate points may have changed. However, one finds that
\be
det( \tilde{C}_{AB} ) = \lambda ^{-4} det( C_{AB}),
\ee
and consequently, both patches have the same conjugate points. Notice 
that (40) is equivalent to the invariance of $det( C_{AB})$  relative
to the
Einstein frame whose metric is  $ g^E _{\mu \nu} = e^{-\phi /2} g_{\mu
\nu}$.

Because $\omega_{-} = 0 $, we now see that the gauge conditions in the NS-NS sector
are preserved by the duality transformations. This property turns out to be
universal for
all the massless fields that may be present. Consider, for example, the YM field
 whose T-duality transformation has been studied in various contexts
\cite{kn:entrop} - \cite{kn:dorn}. For definiteness let us set
\be
{\mathcal A}^{(10)} = {\mathcal A} + \lambda dy  {\mathcal A}_{0},
\ee
and concentrate on the mapping that one gets by gauging of the isometry of the
heterotic
$\sigma$-model action
\cite{kn:entrop}. In this framework $\mathcal A $ transforms as
\be
{\tilde{\mathcal A}}_{0} = {\mathcal A}_{0},\hspace{.5in} {\tilde{\mathcal A}} =
{\mathcal A} +
( {\lambda}^{-1} b - \lambda \omega )  {\mathcal A}_{0},
\ee
and therefore, ${\tilde{\mathcal A}^{(10)}}$ also obeys the gauge condition (13).
 One can check that the
same conclusion is reached when the transformation rule of \cite{kn:ven} or
\cite{kn:dorn} is considered. Hence in all these cases 
the inclusion of the YM Chern-Simons term in $H$ turns out to be of no consequence
for our purpose and the limit
of both 
$\tilde{\mathcal F}$ and $\tilde{H}$ have again the plane wave forms.

     We next cosider the R-R
sector. The transformations of the  R-R fields \cite
{kn:hull}
can be conveniently displayed also by using the 9 +1 decomposition. Following
\cite{kn:stel} we define the D=9 fields 
\be
{F_2}^{(10)} = F_2 + F_1 \wedge (dy + \omega),\hspace{.5in} {F_4}^{(10)} =
F_4 + F_3 \wedge (dy + \omega),
\ee
for the IIA theory. For the IIB theory the correspoding decompositions are
\be
{F_1}^{(10)} = F_1, \hspace{.4in} {F_3}^{(10)} = F_3 + F_2 \wedge (dy + \omega),
\hspace{.4in} {F_5}^{(10)} = F_5 + F_4 \wedge (dy + \omega).
\ee
 In terms of the D=9 quantities the T-duality rules which map the IIA theory into
IIB theory are then
\be
\tilde{F}_1 = - F_1,\hspace{.4in} \tilde{F}_2 = F_2, \hspace{.4in} \tilde{F}_3 = - 
 F_3,
\hspace{.4in} \tilde{F}_4 = F_4,
\ee
together with the rule that $ \tilde{F}_5$ is the D=9 dual of $\tilde{F}_4$.
Since 
these rules only involve $\omega$ and the R-R field strengths of the IIA theory
that one started with, it is obvious
that the gauge choices are preseved also within 
the
R-R sector. Notice that one can also infer the scaling rules
(16)-(18) for
the
IIB fields from those of the IIA theory by invoking the duality (45).

 It therefore
follows that
if one starts with a set of IIA fields and finds the Penrose limit, then the limit
of
the dual set of fields in the IIB theory will be simply the dual of the plane waves
obtained in the IIA theory. Because the two Killing coordinates need not be the
same, one needs two different D
=10 harmonic 
coordinates to describe such a dual pair of plane waves.  One can, of course, always
use the same D=9 harmonic
coordinates on both of the solutions. We shall display the general forms of a IIA -
IIB dual pair in such coordinates. For example, without any loss of generality the 
metric that one obtains from (9) for the IIA theory can
be brought to the form
\be
ds^2 = 2 dudv - h_{ij}(u) x^i x^j du^2 - \delta_{ij} dx^i dx^j - (da - \gamma du
)^2,
\ee
where $ a $ is a new Killing coordinate: $K^\mu = - \lambda {\delta^{\mu}}_a $ and
$\gamma $ is a function which depends linearly on all the transverse
coordinates $x^A = \{x^j, a\}$:
\be
\gamma = {\gamma_A}(u) x^A.
\ee
Here $\gamma_A (u) $ is completely characterized by the norm and the
twist of the Killing one-form 
$ K = {K_\mu} dx^ \mu $. Noting that $ K \wedge dK = {\lambda^2} K \wedge
d\omega$ still
holds after  $\Omega \rightarrow 0$
and writing $ d \omega = {\dot{\omega}_j}(u) du \wedge dx^j $  in the harmonic
coordinates, one
gets
\be
\gamma_A = \{\lambda \dot{\omega}_j, {\dot{\lambda}}/ \lambda \}.
\ee  
The dual of this metric, for example, in the IIB theory is obtained simply by
dualizing the Killing coordinate and $\gamma$:
\be
d\tilde{s}^2 = 2du dv - h_{ij}(u)x^i x^j du^2 -\delta_{ij}dx^i dx^j - (d\tilde{a} -
\tilde{\gamma} du )^2,
\ee
where \be  \tilde{\gamma}_A = \{\lambda^{-1}\dot{b}_j(u), -\dot{\lambda} /
\lambda\},\ee
and $\dot{b}_j$ are defined by $ db =
\dot{b}_{j}(u)du \wedge dx^j$. The dual NS-NS three-forms are given by
\be
H = \frac{1}{2} p_{jk}(u)du \wedge dx^j \wedge dx^k + \lambda^{-1}\dot{b}_j(u) du
\wedge da
\wedge
dx^j,\ee  \be \tilde{H} = \frac{1}{2}p_{jk}(u)du \wedge dx^j \wedge dx^k +
\lambda \dot{\omega}_j(u) du \wedge d\tilde{a} \wedge dx^j, \ee where ${p_{jk}}(u)$
are arbitrary functions. The dilaton $\phi$ is again an arbitrary function of $u$
and $\tilde{\phi} = \phi - \ln \lambda$.

The R-R field strengths also have a similar structure. Provided $p \geq 2 $, a 
D = 10 R-R  p-form 
field strength has the form
\be
F_p = \frac{1}{(p-1) !} {{f^{(p)}}}_{j_1 \ldots j_{p-1}}du \wedge dx^{j_1} \wedge
\ldots
\wedge dx^{j_{p-1}} + \frac{1}{(p-2) !} {{k^{(p)}}}_{j_1 \ldots j_{p-2}}du \wedge
da
\wedge dx^{j_1} \wedge \ldots \wedge dx^{j_{p-2}},
\ee
with arbitrary amplitudes $f^{(p)}_{j_1 \ldots j_{p-1}}(u)$ and $k^{(p)}_{j_1 
 \ldots j_{p-2}}(u)$, and the remaining case is simply: $ F_1 = f^{(1)} du$  where 
$ f^{(1)}(u) = \dot{A}_{0}(u)$. In this notation a set  type IIB R-R plane wave
fields will be the T-dual of its  IIA counterpart if
\be
\tilde{f}^{(1)}(u) = - k^{(2)}(u),\hspace{.2in} {\tilde{k}^{(3)}}_{j}(u) =
{f^{(2)}}_{j}(u),
\hspace{.2in} {\tilde{f}^{(3)}}_{jk}(u) = - {k^{(4)}}_{jk}(u), \hspace{.2in}
{\tilde{k}^{(5)}}_{jkl}(u) = {f^{(4)}}_{jkl}(u).
\ee
In order to satisfy the self-duality condition (7), the remaining D = 9 field
of the IIB theory must obey
\be
{\tilde{f}^{(5)}}_{jklm} = \frac{1}{3 !}{\epsilon_{jklm}}^{noi}
{\tilde{k}^{(5)}}_{noi},
\ee
which means that it is the dual of ${\tilde{k}^{(5)}}_{ijk}$ in the D=7 flat
transverse
space.

 A characteristics of the plane waves is the vanishing of all the
scalar invariants that one can construct from the field strengths and this
is, of
course, shared by all the above field strengths. Moreover, all the field
strengths are
both closed and co-closed under exterior differentiation. The exterior product of a
field strength with another field strength or its Hodge dual is always zero.
Due
these properties, all the field equations of IIA or IIB theory, excepting the
$G_{uu}$ component of the Einstein equations are automatically satisfied.
For example, in the IIA theory the only implication of the field equations is that
\be
h_{jj} = 2 \ddot{\phi} - \frac{\ddot{\lambda}}{\lambda} - \frac{\lambda^2}{2}
{\dot{\omega}}_j {\dot{ \omega}}_j - \frac{1}{2\lambda^2}\dot{b}_j \dot{b}_j -
\frac{1}{4} p_{jk}p_{jk}   - \frac{1}{2} e^{2\phi} [Ramond], \ee where $Ramond$
denotes the
contributions of the R-R sector:  \be Ramond = (k^{(2)})^2 +
{f^{(2)}}_j{f^{(2)}}_j + \frac{1}{2} {k^{(4)}}_{ij} {k^{(4)}}_{ij} + \frac{1}{6}
{f^{(4)}}_{jkl} {f^{(4)}}_{jkl},
 \ee
and this fixes the trace of $h_{jk}(u)$ in terms of the
other fields. Here the trace and the other sums refer to the metric $ \delta_{jk}$
on the D=7 flat
transverse space.

 There are two points worth noting in (56). First, one can infer from (56) that
the most general type IIA (or
IIB)  plane 
wave solution involves a total of 128 arbitrary functions. Half of these always 
belong to 
the NS-NS sector and the remaining 64 functions come from the R-R sector. These
numbers are precisely the numbers of degrees of freedom of the massless
states in
the first quantized type II string theories. 
Secondly,
(56) shows that, with the above assumptions about the isometry, plane waves
constitute 
a T-duality invariant family. This result was already known \cite{kn:kl} within the
NS-NS sector  where $ h_{jk}$ and $ p_{jk}$ are inert to duality and the dual
roles of $ \phi$ and $\lambda$ as well as of $ {\dot{\omega}}_j$ and $\dot{b}_j$ are
manifest. Because (54) and (55) hold, the duality invariance of the R-R sector is
also now
manifest.

Some particular choices of the arbitrary functions appearing in (56) lead to
interesting
generalizations of 
the well known solutions. One such class is the case of the sandwich waves
\cite{kn:gib} where all the amplitutes are taken to be non-zero only over a
finite interval of $u$. This leads to a spacetime in which two flat regions are
connected by plane waves of finite duration. It would be interesting to see how the
boundary state formalism for the D-branes can be applied on such a geometry. 
Another class is obtained by choosing
all the amplitudes to be constants so that $u$ is also  a global Killing
coordinate and $h_{jk} = c \delta_{jk}$, where $c$ is a constant determined by (56).
With these assumptions one gets a generalization
of
the Nappi-Witten solution \cite{kn:nappi} to D=10 and to a non-zero R-R sector. In
this
case
the NS-NS sector of the solution corresponds to a WZNW model based on the
ten-dimensional Heisenberg group \cite{kn:keh}.

Notice that although a spacelike isometry is initially an additional assumption,
 this does not entail an additional symmetry on the spacetime
that one gets through the limit.
Viewed after the limit, the Killing vector $K^\mu$ 
is simply a member of the 17-parameter group of motions of the D= 10 plane wave
spacetimes. If
the orbits
of $K^\mu$ are assumed to be not compact, which is the case of the usual plane
waves, 
$M_{10}$ as well as its dual has the standard
$ R^{10}$ topology after the limit. In this case duality is just a mapping
between two different sets of solutions.  When $K^\mu$ has compact orbits so that
the quantum equivalence of the underlying string theories can be considered, the
manifold  $M_{10}$ that one obtains by the limit has the $ R^9 \times S^1$
topology and moreover, dualization does not
bring in a
twist at the field theory level.
In other words, after the limit the dual manifold has again the $  R^9 \times
S^1$ topology. Since the topological properties are not hereditary properties in the
sense of \cite{kn:ger}, a particular topology for $ M_{10}$ need not be assumed
prior to the limit in the non-compact case. In the case of a compact isometry, one
must start with a $M_{10} =  M_9 \times S^1$, where the topolgy of $M_9$ is
initially unspecified but $ y \equiv y + 2 \pi \sqrt{\alpha '} R_b$ on $S^1$ .
Letting 
$ y = \Omega x $ and noting that $\alpha '$ also scales, this implies $ x \equiv x
+ 2 \pi \sqrt{\bar{\alpha} '} R_b$ on the plane wave Killing coordinate $x$. 
In the coordinate system of (46) this identification corresponds to a ``local
compactification": $ a \equiv a + 2 \pi \sqrt{\bar{\alpha} '} \lambda (u) R_b $.

In type I and heterotic theories plane waves are known to be exact
solutions
which preserve half of the Poincar\'e supersymmetry
\cite{kn:guv},\cite{kn:ama}, \cite{kn:hor}.
 These plane waves therefore satisfy the field equations not only at the leading
order 
but in all orders of $\alpha '$. Moreover, their behavior under the T-duality is not
affected by the higher
order
$\alpha '$  corrections to the Buscher rules \cite{kn:entrop}. When the self-dual
five-form is switched off, type IIB plane waves are also exact solutions
\cite{kn:seng}. The plane waves
of the IIA theory, on the other hand, are known to admit at least chiral Killing
spinors which
preserve again 1/2
supersymmetry \cite{kn:spin}. These Killing spinors do not depend on the
Killing coordinate used in duality and consequently, one can conclude that the plane
wave duals in the IIB theory
are also supersymmetric \cite{kn:kal}. Since the solutions on which the limit is
applied are not necessarily exact, supersymmetric or have exact T-duals, it is clear
that the absence of these basic properties are not hereditary. The presence of any
one
of
these properties in an
initial configuration is, of course, hereditary.

\section*{Appendix}
Our conventions are as follows: In all $D \geq 2$ we use the ``mostly
minus''
signature $ (+,-, \ldots, -)$ and the orientation $ \epsilon_{012 \ldots
D-1} = 1$. The Ricci tensor is defined as $R_{\mu \nu} = {R^\lambda}_{\mu
\nu
\lambda}$ and the Riemann curvature obeys $ (\nabla_\nu \nabla_ \mu -
\nabla_\mu \nabla_\nu) T_\kappa = {R^\lambda}_{\kappa \mu \nu} T_\lambda $
for an arbitrary $ T_\mu$. The Hodge dual of a p-form $(p \leq D)$ is
defined
by
\begin{eqnarray}
*( V^{\alpha_1} \wedge \ldots \wedge V^{\alpha_p}) =
\frac{(-1)^{(D-1)}}{(D-p) !}
   \epsilon^{\alpha_1 \ldots \alpha_ p \alpha_{p+1} \ldots \alpha_ D}
 V_{\alpha_{p+1}} \wedge \ldots \wedge V_{\alpha_ D}, \nonumber 
\end{eqnarray}
in terms of an orthonormal basis $\{ V^\alpha \}$. 
   
\section*{Acknowledgements}

I thank The Abdus Salam International Centre for
Theoretical Physics
for the warm hospitality. The research reported in this paper has been
supported in part by the Bogazi\c{c}i University Foundation (B{\" U}VAK)
and the
Turkish
Academy of Sciences(T{\"U}BA).

\end{document}